\documentclass[mathematics,article,accept,pdftex,oneauthor]{Definitions/mdpi} 

\firstpage{1} 
\pubvolume{1}
\issuenum{1}
\articlenumber{0}
\pubyear{2024}
\copyrightyear{2024}
\externaleditor{Academic Editors: Francisco G. Montoya and Alfredo Alcayde}
\datereceived{18 May 2024} 
\daterevised{18 June 2024} 
\dateaccepted{20 June 2024} 
\datepublished{ }
\hreflink{https://doi.org/} 

\Title{Quaternion Spin}

\TitleCitation{Quaternion Spin}

\Author{{Bryan Sanctuary} 
 $^\dagger$\orcidA{}}

\AuthorNames{Bryan Sanctuary}

\AuthorCitation{Sanctuary, B.}

\address[1]{\hl{Department of} Chemistry, McGill University,  
{Montreal, QC H3A 0B8}, Canada; bryan.sanctuary@mcgill.ca}

\firstnote{\hangafter=1 \hangindent=1.05em \hspace{-0.82em}Retired Professor.}

\abstract{We present an analysis of the Dirac equation when the spin symmetry is changed from SU(2) to the quaternion group, $Q_8$, achieved by multiplying one of the gamma matrices by the imaginary number, $i$. The reason for doing this is to introduce a bivector into the spin algebra, which complexifies the Dirac field.  It then separates into two distinct and complementary spaces: one describing polarization and the other coherence. The former describes a 2D structured spin, and the latter its helicity, generated by a unit quaternion.}


\begin{document}
	
	\section{Introduction}
	
	Spin, first observed by Stern and Gerlach~\cite{SG}, reveals two up and down states. Spin is measured to be angular momentum of $\frac{\sqrt 3}{2}\hbar$ magnitude, a~vector quantity, belonging to the SU(2) group. Spin is a fundamental property of Nature, purely quantum with no classical analog. The mathematical basis for spin is the Dirac equation~\cite{6 dirac}. Dirac's analysis introduces his relativistic equation by linearizing the Klein-Gordon equation while respecting the conservation of mass and energy. He was led to his gamma matrices with four states rather than the two measured~\cite{PerSch}. He surmised that his equation described two spins rather than one. The~two are mirror-image twins of each other, which Dirac interpreted as a matter-antimatter pair~\cite{6 dirac}. From~this hole theory, antimatter production and~the sea of electrons model followed~\cite{hole}.
	
\begin{figure}[H]
		
		\hspace{90pt}	\textbf{{A spin} $\frac{1}{2}$ in free flight}
		
		\includegraphics[width=0.7\linewidth]{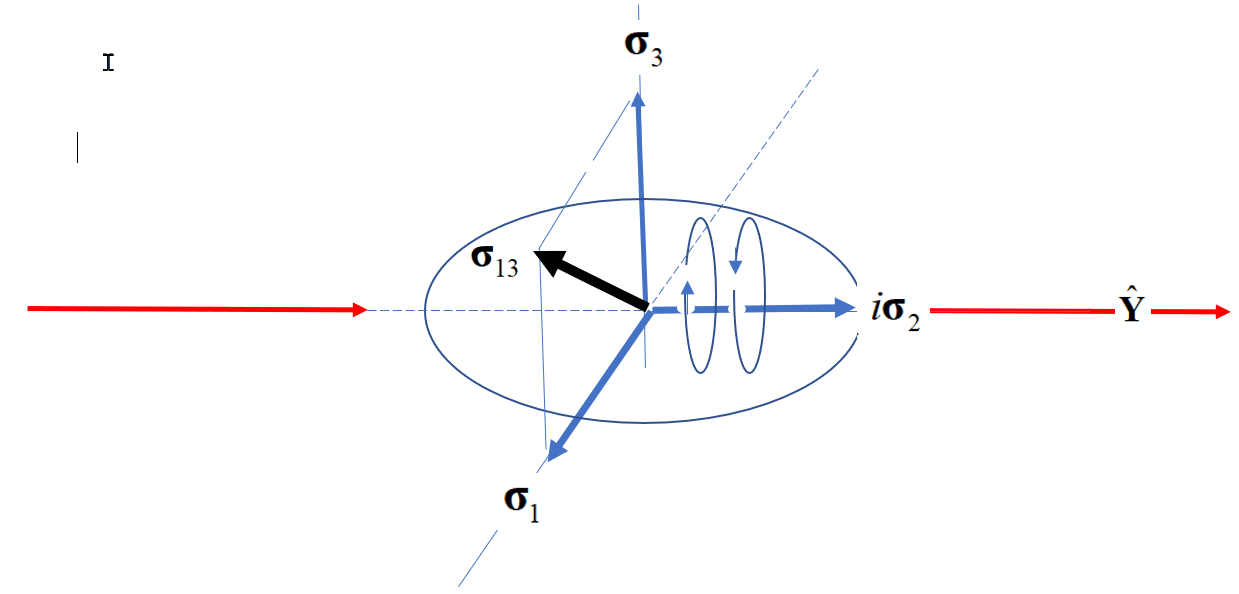}		
		\caption {Two properties of spin: two orthogonal spin-$\frac{1}{2}$ polarization vectors, $\sigma_{1}$ and $\sigma_{3}$, perpendicular to the direction of linear momentum, $Y$. The~helicity, generated by $i\sigma_2$, is in the direction of propagation, $e_2=Y$, and~spins R or L. The~two spin-$\frac{1}{2}$  vectors, $\sigma_3$ and $\sigma_1$, couple to give a composite boson, $e_{13}$, of~magnitude 1. These properties show that the spin is geometrically equivalent to a photon. 
		}	
		\label{fig:freespin}
	\end{figure}

	Under the quaternion group, the~two point particles that Dirac found are replaced by one structured particle called quaternion spin, or~Q-spin, that carries two complementary properties: polarization and coherence. The~coherence is helicity, which spins the axis of linear momentum in free flight, as shown in Figure~\ref{fig:freespin}, giving the two helicity states of L and R. In~addition, two mirror states~\cite{mirror} emerge which describe two orthogonal magnetic axes, each with a magnetic moment of $\mu$. Each is also perpendicular to the axis of linear momentum.  The figure shows that Q-spin is geometrically equivalent to a photon.  The~two magnetic fermionic axes $e_3$ and $e_1$ each carry a spin-$\frac{1}{2}$, which can couple to give a composite spin-1 boson, $e_{13}$.  All three axes are orthogonal to $e_2=Y$, the axis of linear momentum, Figure~\ref{fig:freespin}. Structured spin makes the intrinsic angular momentum of Dirac spin extrinsic.
	
	To motivate this {discussion}, consider the well know equation for the geometric product of Pauli spin components, 
	\begin{equation} \label{dyadic}
		{{\sigma }_{i}}{{\sigma }_{j}}={{\delta }_{ij}}+{{\varepsilon }_{ijk}}i{{\sigma }_{k}}
	\end{equation}
	Arising from Geometric Algebra~\cite{GA,GABiv}, the first term describes a symmetric component that gives rise to polarization and measured Dirac spin. The second term is anti-symmetric and depends upon a bivector, $i\sigma_k$, and the Levi Civita third-rank anti-symmetric tensor. Since $i$ cannot simultaneously be equal and not equal to $j$, the~geometric product, Equation~(\ref{dyadic}), is complementary.  There is, however, no bivector in the Dirac equation.  We introduce a bivector by multiplying a gamma matrix by the imaginary number, $\tilde{\gamma}^2_s \equiv  i\gamma^2_s$. This makes the Dirac field complex, which is the origin of helicity.  This paper aims to include this anti-symmetric term as a property of spin, even though it is not measurable.

	The procedure here has similarities to Penrose's Twistor theory~\cite{penrose1,penrose2}.  Essentially, Twistor theory complexifies Minkowski space, a~four-dimensional real manifold $M$, into complex Twistor space, $T$. As~a complex space, it has two projections into helicity states of $+$ and $-$, denoted by $PT^{\pm}$. The~boundary between the two, $PN$, is a real space of null vectors, which are light rays from the light cone.  Different slices in $T$ lead to different projections, groups, and~algebras.
	
	Here, we complexify the Dirac field in spin spacetime, not Minkowski space. No additional parameters are introduced, and a non-Hermitian Dirac equation determines the field. Under~parity, this splits further into complementary spaces. In the following, these ideas are formalized.

	This paper is the first of several in which the properties and foundations of Q-spin are presented. A~second paper, “Spin helicity”~\cite{sanctuary2}, discusses the geometry and conservation of correlation. A~third paper~\cite{sanctuary3} studies and simulates the EPR correlation between a pair of spins. A~general summary~\cite{sanctuary4} describes some consequences of this symmetry~change.
	
	This four-state spin is called Q-spin to distinguish it from the usual Dirac spin. A modified form of the Dirac equation admits both polarization and helicity, complementary elements of reality. 
	It describes spin as it exists in the absence of interactions and, therefore, in free flight. Additionally, when measured, the~boson spin decouples into a fermion of spin-$\frac{1}{2}$. This is the measured spin that Dirac formulation.  
	
	Q-spin is a boson of odd parity in free flight, a~wave, and a fermion of even parity when measured, a~particle.  A similar conclusion was reached by Geurdes, \cite{geurds2}, who showed the Maxwell field equations for many photons is related to the Dirac equation describing fermions. Spin has also been studied using quaternions as bosons and fermions~\cite{adler,das} and extends to relativistic quantum mechanics, as shown by Adler~\cite{adler}, Rotelli~\cite{rotelli}, Colladay et al.~\cite{colladay}, and Leo et al.~\cite{rotelli1,rotelli2}, who also reformulated the Dirac equation from quaternions in an electromagnetic field~\cite{Bhatt,Rawat2,Gsponer}. A large volume of work studies the Supersymmetric formulation of spin using quaternions, Davies~\cite{Davies2} 
	
	Our work has similarities to the above. We emphasize the structure of Q-spin as a consequence of complexification. We also present the fermion and boson forms of Q-spin and describe mechanisms for a transition between the two.

	\section{Spin Spacetime~Algebra}		  
	
	No bivector is found in the Dirac equation because his point-particle spin is defined in Minkowski space, a four-dimensional real manifold. In contrast, introducing a bivector gives spin structure and~since it can be oriented randomly relative to Minkowski space; we introduce spin spacetime, $\left( \beta_s,e_1,e_2,e_3 \right)$, which is the Body-Fixed (BFF) of one spin. Minkowski space is the Laboratory-Fixed Frame (LFF) $\left( \beta,X,Y,Z \right)$.  
	
	Dirac's gamma matrices, $\left( {{\gamma }^{0}},{{\gamma }^{1}},{{\gamma }^{2}},{{\gamma }^{3}} \right)$, represent the 4x4 Dirac field. Within~this field, there are two point-particle spins of $\frac{1}{2}$---each described by the three Pauli spin components and the identity ($I,\sigma_{X},\sigma_{Y}, \sigma_{Z})$---which belong to the SU(2) group; each is the mirror twin of the other: Dirac's matter--antimatter~pair. 
	
	Introducing $\tilde{\gamma}^{2}_s =i\gamma^{2}_s$  complexifies~spin spacetime (subscript $s$) \cite{penrose1,penrose2}. 
	Minkowski space obeys Clifford algebra $\mathbb{C}{{\ell }_{1,3}}$.  In~contrast, the~Clifford algebra of spin spacetime is $\mathbb{C}{{\ell }_{2,2}}$. An extensive body of literature discusses this split group~\cite{penrose2,Jain,Atiyah,Harnad}.  The~immediate consequence of introducing the bivector is that the Dirac equation becomes non-hermitian, with two fields expressed by $\left( {{\gamma }_s^{0}},{{\gamma }_s^{1}},\pm{{\tilde{\gamma} _s^{2}}},{{\gamma }_s^{3}} \right)$, and the~solutions are mirror states, $\psi^\pm$ with no parity. The~$\pm$ division is complex~conjugation.
	
\begin{figure}[H]
		\hspace{70pt}	 \textbf{Minkowski and spin spacetime}\\
		\includegraphics[width=0.8\linewidth]{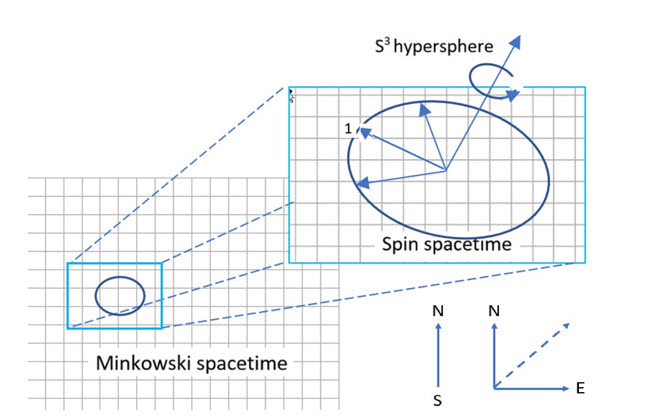}
		\caption{Spin is oriented in spin spacetime by the BFF basis vectors $(e_1,e_2,e_3)$, which spin about the axis $e_2$ so that in Minkowski space, with~the basis vectors $(X,Y,Z)$, only a smeared-out image of the precessing spin is projected. The lower right insert contrasts Dirac spin and Q-spin, which is displayed as the resonance formed from coupling the N and E axes.}
		\label{fig:spinspace}
		
	\end{figure}

	The mirror states can be combined into odd and even parity states, and upon doing this, the~field separates once more into two distinct spaces and the algebra changes from $\mathbb{C}{{\ell }_{2,2}}$ to~a 2D plane with the algebra $\mathbb{C}{{\ell }_{1,2}}$, and a connection to the $S^3$ hypersphere via $\tilde{\gamma} _s^{2}$.  Spin spacetime decomposes into two complementary spaces: polarization spacetime (0,1,3) of even parity and coherence space (2) of odd parity. It has the structure of a 2D plane of polarization, Figure~\ref{fig:spinspace}.

	The bivector, $i\sigma _2$, connects spin spacetime to the complementary space of the helicity states generated by quaternions in the $S^3$ hypersphere. This has four spatial dimensions and cannot be measured. Its only role is to spin the axis of linear momentum, $Y=e_2$, either L or R, which are the two helicity states. Note that the helicity generates its own $S^3$ hyperspace in free flight. It does not exist when measured. 
	
	Q-spin is one particle with four states, not two particles with two states each. This is an entirely different interpretation from Dirac's matter-antimatter pair. It suggests that only Q-spin electrons and no positrons formed in the Big Bang. This obviates the need for Baryon asymmetry, \cite{Baryon}, to explain the dominance of matter.  We do not deny that antimatter is produced~\cite{Ahmadi}, but~not as Dirac proposed if quaternion spin is accepted. Instead, consistent with observations, antimatter is produced in small amounts from radioactive isotopes and pair particle production from high-energy photons. There can be no mirror universe under quaternion symmetry.
	
	The treatment here shows that the two fermionic axes are exact reflections of each other, remaining in phase with equal and opposite frequency. 
	Q-spin does not have the negative energy problem Dirac~encountered.
	
	When the spin is measured, the helicity stops, and the usual two polarized states of up and down are observed in some direction,   $\left| \pm ,\mathbf{\hat{n}} \right\rangle $.  Away from polarizing fields, the spinning axis of linear momentum averages out the spin polarization. Only the helicity is then~present. 
	
	\textls[-15]{The spin-polarized structure can be expressed in Minkowski space. The~bivector~cannot. }
	
	\section{Mirror States and~Parity}	
	The spin space-time gamma matrices  ${\left( \gamma _{s}^{0},\gamma _{s}^{1},\pm\tilde{\gamma}_{s}^{2},\gamma _{s}^{3} \right)}$ anticommute and have a different signature from Minkowski space,
	\begin{equation}                  
		\tilde{\eta}^{\mu\nu}_s=\left(
		\begin{array}
			[c]{cccc}
			+1 & 0 & 0 & 0\\
			0 & -1 & 0 & 0\\
			0 & 0 & +1 & 0\\
			0 & 0 & 0 & -1
		\end{array}
		\right)
	\end{equation}
	so the term  $\tilde{\gamma}^{2}_s$ is not a spatial component, but~rather, time-like and a~frequency. 
	
	The commutation relations are changed from the usual three-dimensional generator of rotations in Minkowski space,
	\begin{equation}\label{rotation}
		S_{(3)}^{ij}=\frac{i}{4}\left[  \gamma^{i},\gamma^{j}\right]  =\frac{1}
		{2}\varepsilon_{ijk}\sigma_{k}I_{4}		
	\end{equation}
	to ones that generate rotations in only two dimensions in spin {spacetime} 
	\begin{equation}
		S_{(2)}^{ij}    =\frac{i}{4}\left[  \gamma^{i}_s,\gamma^{j}_s\right]  =\frac{i}{2}\varepsilon_{i2j}\tilde{\sigma}_{s2}I_{4}
	\end{equation}
	\begin{equation}
		S_{(2)}^{i2}    =\frac{i}{4}\left[  \gamma^{i}_s,\tilde{\gamma}^{2}_s\right]=\frac{i}{2}\varepsilon_{i2j}\sigma_{sj}I_{4} \label{damps}
	\end{equation}
	{The} 
 former equation describes rotations in the $31$ plane in the direction of $2$, whereas the imaginary term in the latter equation damps all rotation attempts out of the $31$ plane. This leaves a disc of spin polarization.

\begin{figure}[H] 
		\hspace{100pt}	\textbf{Mirror states of Q-spin}\\
		\includegraphics[width=0.8\linewidth]{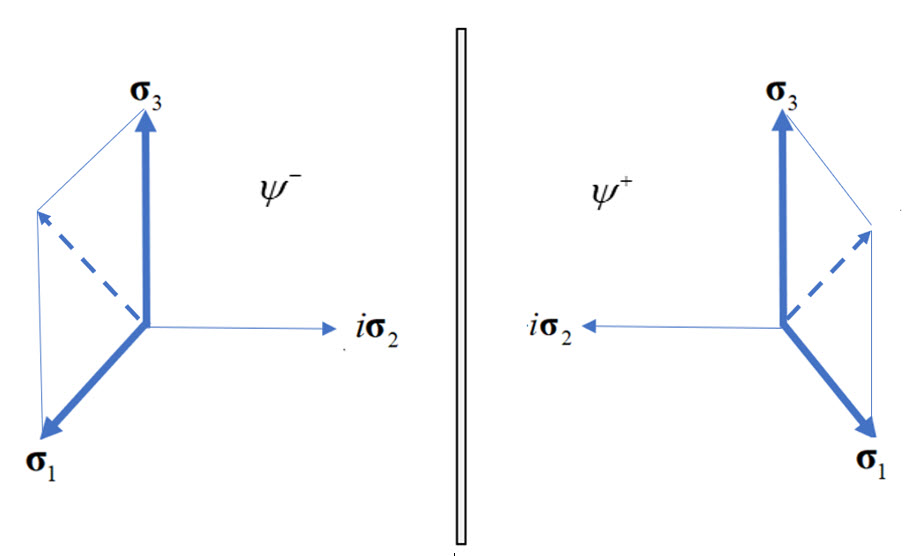}	
		\caption {The mirror states of a Q-spin with $\psi^{+}$ on the right and $\psi^{-}$ on the left. Note that adding these states is independent of $i\sigma_2$\ and subtracting them is independent of $\sigma_1$ and $\sigma_3$.}
		\label{fig:mirror}
		\end{figure}

	Two new equations in spin spacetime follow from the gamma algebra, which gives a non-Hermitian equation due to  $\tilde{\gamma}^{2}_s$,
	\begin{equation}	\label{Eq6}
		\left( i\gamma _{s}^{0}{{\partial }_{0}}-i\gamma _{s}^{1}{{\partial }_{1}}\pm i\tilde{\gamma }_{s}^{2}{{\partial }_{2}}-i\gamma _{s}^{3}{{\partial }_{3}}-m \right){{\psi }^{\pm }}=0 	
	\end{equation}
	We suppress the subscript $s$ on the derivatives. 
	By treating a spin in free flight in an isotropic environment, the two axes $(1,3)$ are indistinguishable. Therefore, permutation with the parity operator $P_{13}$ does not change the $(1,3)$ dependence in Equation~(\ref{Eq6}), but~the bivector $i\sigma_2=\sigma_3\sigma_1$ is anti-symmetric to the \mbox{13 permutation}. Therefore, the above equations provide two solutions in the left- and right-handed coordinate frames, which are mirror states~\cite{mirror,mirror2}, Figure~\ref{fig:mirror}:
	\begin{equation}\label{mirrostates} P_{13}\psi^{\pm}=\psi^{\mp}
	\end{equation}
	The~parity operator is given by~\cite{kaempffer},
	\begin{equation}
		{{P}_{13}}=\frac{1}{2}\left( {{I}^{1}}\otimes {{I}^{3}}+{{\sigma }^{1}}\cdot {{\sigma }^{3}} \right)
	\end{equation}
	and permutes the  (1,3) labels,
	\begin{equation}
		{{P}_{13}}{{\sigma }^{1}}P_{13}^{-1}={{\sigma }^{3}}\text{ and }{{P}_{13}}{{\sigma }^{3}}P_{13}^{-1}={{\sigma }^{1}}
	\end{equation}
	The anti-commutation of the $\gamma^{\mu}_s$ matrices ensures energy is conserved and the Klein--Gordon equation is~recovered. 
		Adding and subtracting the two equations in Equation~(\ref{Eq6}) leads to their separation into a Hermitian part and an anti-Hermitian {part,} 
	\begin{equation}
		\left(  i\gamma'^{0}_s\partial_{0}-i\gamma^{1}_s\partial_{1}-i\gamma^{3}_s\partial_{3}-m\right)  \Psi^{+}  =0\label{nonH1}
	\end{equation}
	\begin{equation}\label{nonH2}
		\tilde{\gamma}^{2}_s\partial_{2}\Psi^{-}   =0
	\end{equation}
	where the two mirror states combine into states with odd and even parity, $P_{13}\Psi^{\pm}=\pm\Psi^{\pm}$, with the definition
	\begin{equation} \label{parity}
		\Psi^{\pm}=\frac{1}{\sqrt{2}}\left(  \psi^{+}\pm\psi^{-}\right)
	\end{equation}
	The even-parity states describe polarization, and the odd-parity states describe its~helicity. 
	
\begin{figure}[H]
		
		\hspace{150pt}		\textbf{Spin decoupling}\\
		\includegraphics[width=1\linewidth]{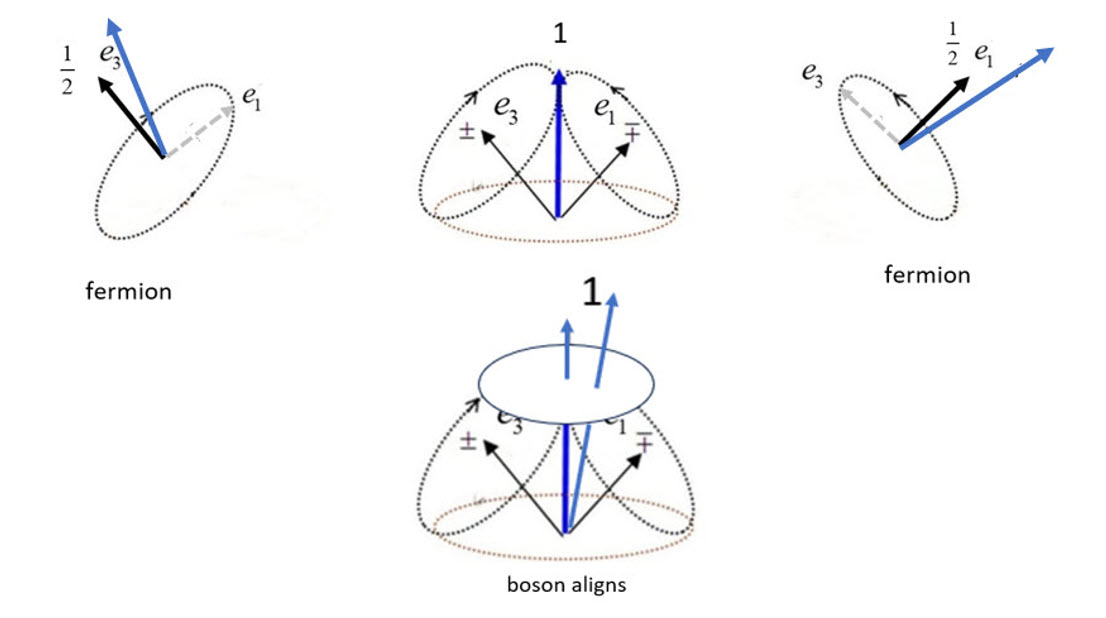}
		\caption{The~longer arrow denotes the direction of the polarizing field. \textbf{Middle top:} The two mirror states are in free flight, and $e_1$ and $e_3$ couple to give a boson spin-1. \textbf{Left and right:} The fermionic axis closer to the field axis aligns, and~the boson decouples. \textbf{{Middle} 
				bottom:} When the boson spin is close to the field, it initially precesses as a spin-1 without decoupling. }
		
		\label{fig:Rqspin}  	
	\end{figure}

	The separation of the Dirac field into reflective states means each axis precesses in the opposite direction. These two polarization axes, each with a magnetic moment $\mu$, constructively interfere, producing resonance and~purely coherent spin, as shown in Figure~\ref{fig:Rqspin} (middle top). Such a resonance structure lowers the energy and stabilizes the 2D structure over that of two point-particle spins.

	The Hermitian part, Equation~(\ref{nonH1}), is the same as the usual Dirac equation but in two dimensions rather than three. It describes a disk, as~visualized in Figures~\ref{fig:freespin} and \ref{fig:spinspace}.
	
	The bivector component, (2), describes a massless Weyl spinor in coherent space,  {Equation~(\ref{nonH2})}.
	Time does not exist within this space beyond the constant frequency of its spinning. Time and rest mass remain in polarization space. Similar to the two complementary inverse spaces of position and momentum, the two spin spaces carry the two complementary properties of polarization and coherence.
	
	We redefine the spinor mirror states as ${{\psi }^{+}}\equiv {{\psi }_{R}}$ and ${{\psi }^{-}}\equiv {{\psi }_{L}}$.
	\section{The Weyl~Spinor} 
	From  {Equation~(\ref{nonH2})} 
	and using references~\cite{PerSch,massive}, a Weyl spinor transforms under boosts and {rotations as follows:} 
	\begin{equation}
		{{\psi }_{R}}\to \left( 1-i\chi \frac{{{\sigma }_{2}}}{2}+i\beta  \frac{{{\sigma}_{2}}}{2} \right){{\psi }_{R}}\left( 0 \right) 
	\end{equation}
	\begin{equation}
		{{\psi }_{L}}\to \left( 1-i\chi \frac{{{\sigma}_{2}}}{2}-i\beta \frac{{{\sigma}_{2}}}{2} \right){{\psi }_{L}}\left( 0 \right)  
	\end{equation}
	Since time exists only in polarization space, Equation~(\ref{nonH1}), a~boost of polarizations carries along the spinors.  There are no boosts in coherent space; the left and right wave functions are equal.
	\begin{equation}
		{{\psi }_{R}}={{\psi }_{L}}
	\end{equation}
	This state is a unit quaternion which spins the axis of linear momentum in coherence space (2)
	by angle $\chi$, thereby generating helicity.
	\begin{equation} \label{quat}
		{{\psi }_{L}(\chi)}=\exp \left( -i\frac{\chi 	}{2}{{\sigma }_{2}} \right){{\psi }_{L}}\left( 0 \right)=\left( \cos \frac{\chi }{2}-i{{\sigma }_{2}}\sin \frac{\chi }{2} \right){{\psi }_{L}}\left( 0 \right)
	\end{equation}
	
	The usual definition~\cite{griffin} identifies helicity as the projection of the spin vector onto the axis of linear momentum in Minkowski space. The~helicity of the axis, L or R, gives a spin state of $+1$ or $-1$. Spin and helicity are related and not independent. Q-spin is quite different~. 
	
	Here, helicity is defined only in quaternion space, the~$S^3$ hypersphere~\cite{hyper,joy1,joy2}, where there is no momentum to contract.  Choosing $Y= e_2$ connects Minkowski space to spin spacetime and finally to the $S^3$ hypersphere, where the component $i\sigma_2$ generates the quaternion in Equation~(\ref{quat}), and~provides a mechanism  for~helicity. 
	
	Within the spinning disc in the (3,1) plane, the fermionic axes couple to give the composite boson of spin-1.  However, the~rapid spinning averages out the boson polarization in the disc, so only helicity is present in free flight: an electron is then a boson of odd parity, $e^{-}_B$. 
	
	Helicity is a distinct element of reality and complementary to observed polarized spin. All we observe of the helicity in our spacetime is its stereographic projection, which is the spinning of the $Y$ axis, giving a spinning disc of angular momentum in Minkowski space, Figure~\ref{fig:spinspace}. 
	
	\section{The 2D Dirac~Equation} 
	We can transform from the BFF of the spin to the LFF using
	\begin{equation}\label{BFF}
		\begin{aligned}
			& {{e}_{3}}=\cos \theta Z+\sin \theta \cos \phi X+\sin \theta \sin \phi Y \\ 
			& {{e}_{1}}=-\sin \theta Z+\cos \theta \cos \phi X+\cos \theta \sin \phi Y \\ 
			& {{e}_{0}}=\cos \phi Y-\sin \phi X \\ 
		\end{aligned}
	\end{equation}
	giving
	\begin{equation}
		\begin{aligned}
			& \gamma _{s}^{1}=\left( -\sin \theta {{\gamma }^{3}}+\cos \theta \cos \phi {{\gamma }^{1}}+\cos \theta \sin \phi {{{{\gamma }}^{2}}} \right) \\ 
			& \gamma _{s}^{3}=\left( \cos \theta {{\gamma }^{3}}+\sin \theta \cos \phi {{\gamma }^{1}}+\sin \theta \sin \phi {{{{\gamma }}^{2}}} \right) \\ 
			& {{p}_{3}}=\mathbf{p}\cdot {{\mathbf{e}}_{3}}=\left( \cos \theta {{p}_{Z}}+\sin \theta \cos \phi {{p}_{X}}+\sin \theta \sin \phi {{p}_{Y}} \right) \\ 
			& {{p}_{1}}=\mathbf{p}\cdot {{\mathbf{e}}_{1}}=\left( -\sin \theta {{p}_{Z}}+\cos \theta \cos \phi {{p}_{X}}+\cos \theta \sin \phi {{p}_{Y}} \right)  
		\end{aligned}
	\end{equation}
	The following expression is independent of $\theta$:
	\begin{equation}
		\begin{aligned} \label{project1}
			\gamma _{s}^{1}{{p}_{1}}+\gamma _{s}^{3}{{p}_{3}}= {{\gamma }^{3}}{{p}_{Z}}+\left( \cos \phi {{\gamma }^{1}}+\sin \phi {{{{\gamma }}^{2}}} \right)\left( \cos \phi {{p}_{X}}+\sin \phi {{p}_{Y}} \right) 
		\end{aligned}
	\end{equation}
	Taking the linear momentum in the direction  $Y=e_2$   requires setting $\phi = 0$, giving,
	\begin{equation} \label{project}
		\gamma _{s}^{1}{{p}_{1}}+\gamma _{s}^{3}{{p}_{3}}={{\gamma }^{1}}{{p}_{X}}+{{\gamma }^{3}}{{p}_{Z}}
	\end{equation}
	The polarization in spin spacetime is projected onto Minkowski space. The~spinning from helicity is in coherent space, which spins the polarization in Minkowski~space.  
	
	In contrast, a two-state fermion spin forms the unusual Dirac point particle  $e^-_F$ when it encounters a polarizing field. A Fermi electron is even to parity. 
	It has two states of up and down.  For~boson electrons, the two polarized states are suppressed, leaving the two helicity states of L and~R.

	We can define a momentum vector, $\mathbf{p}=p_3 e_3 + p_1 e_1$, and the equation for 2D polarization becomes
	\begin{equation}
		\label{KG}
		\left( \begin{matrix}
			E-m & -\mathbf{p}\cdot \sigma  \\
			+\mathbf{p}\cdot\sigma & -\left( E+m \right)  \\
		\end{matrix} \right)\left( \begin{matrix}
			{{u}^{+}}  \\
			{{v}^{+}}  \\
		\end{matrix} \right)=0
	\end{equation}
	where the even-parity state is written as	${{\Psi }^{+}}=\left( \begin{matrix}
		{{u}^{+}}  \\
		{{v}^{+}}  \\
	\end{matrix} \right)$. This leads to the same Klein-Gordon equation in Minkowski and spin {spacetime,} 
	\begin{equation} \label{KG3}
		\left( \partial _{0}^{2}-\partial _{Z}^{2}-\partial _{X}^{2}-{{m}^{2}} \right)\psi =0
	\end{equation}
	\begin{equation}
		\left( {\partial_s}_{0}^{2}-\partial _{3}^{2}-\partial _{1}^{2}-{{m}^{2}} \right)\psi_s =0 
	\end{equation}
	with eigenvalues for the latter of
	\begin{equation} \label{KE}
		{E}=\pm \sqrt{{{m}^{2}}+p_{3}^{2}+p_{1}^{2}}.
	\end{equation}
	We interpret the two energy states as internal energy, which is absent for point particles. This is caused by the precession of the two spin axes on the same particle, Figure~\ref{fig:Rqspin}.  As mirror states, they are depicted as being in phase with equal, but opposite, energy—the two couple to give a resonance spin-1. Precession, as shown, gives one component of, say, $m=+1$.  Reversing these precessions gives the $m=-1$ component.  The~$m=0$ component cannot form since it would violate the reflective symmetry between the mirror states. Note also that a photon has no $m=0$ component.	The two axes form the resonance boson, as shown in Figure~\ref{fig:Rqspin}. Rather than Dirac's matter-antimatter pair, Q-spin resolves the negative energy problem Dirac encountered because the two axes must have equal energy but spin oppositely.
	
	We define the helicity matrix, $H_g={{\gamma }^{1}}{{\tilde{\gamma }}^{2}}{{\gamma }^{3}}$,  which gives the spatial gamma matrices:
	\begin{equation} \label{gammas}
		{{\sigma }_{i}}\otimes H_g={{\sigma }_{i}}\otimes \left( \begin{matrix}
			0 & +I  \\
			-I & 0  \\
		\end{matrix} \right)={{\gamma }^{i}}
	\end{equation}
	The gamma algebra of Q-spin is virtually the same as for Dirac~spin. 
	
	\section{Quaternion~Spin}
	
	In this section, we present more specific equations that describe the structure and some properties of Q-spin. The equations that lead to the illustrations in  Figure~\ref{fig:Rqspin} are given.
	
\begin{figure}[H]
		\hspace{50pt}		\textbf{Q-spin in a field}\\
		\includegraphics[width=0.5\linewidth]{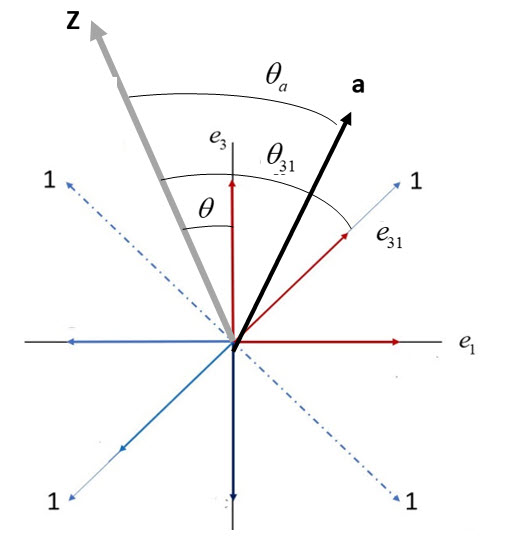}
		\caption{The BFF showing the $\left(e_3,e_1\right)$ plane and the bisectors of the quadrants with boson spins-1. The~$e_{31}$ boson is labeled. The~plane is oriented in the LFF by the $Z$ axis, and~the angle $\theta_{31}$ is shown.  Also, $\theta_a$ orients the field vector $\mathbf{a}$ in the LFF. The~angle $\theta$ is the orientation of a spin vector on the Bloch~sphere.} 		\label{fig:Infield}		
	\end{figure}

	Figure~\ref{fig:Infield} shows the BFF with the $Y$ axis perpendicular to the screen.  The four bisectors are shown, and the first quadrant is $\left(e_3,e_1\right)$. Also shown is the long LFF $Z$ axis oriented relative to the BFF by angle $\theta$. The~field axis, $\mathbf{a}$, is oriented by angle $\theta_a$ from $Z$, and~finally, the~boson spin, $e_{31}$, is at angle $\theta_{31}$ from the $Z$ axis. 	The spinning disc is orthogonal to the direction of motion, and therefore, the polarizing filter and the disc are~co-planar.

	The complementary attributes of spin, polarization and coherence, simultaneously exist, but only one is manifest at any instant. Just as the geometric product, Equation~(\ref{dyadic}), is the sum of two complementary contributions, so too we extend the usual definition of spin,  $\sigma$, to~define Q-spin, $\Sigma_k$, as~possessing both these properties:
	\begin{equation}\label{sigmar}
		\Sigma_k =\sigma_k +{{\underline{\underline{h}}}^k_{g}}=\sigma_k +\underline{\underline{\underline{\varepsilon }}}\cdot i\sigma_k :\left( k=1 \text{ or }3 \right)
	\end{equation}	
	Based on the geometric product, 
	Equation~(\ref{dyadic}), the~geometric helicity operator, ${{\underline{\underline{h}}}_{g}=\underline{\underline{\underline{\varepsilon }}}\cdot i\sigma}$, is an anti-symmetric, anti-Hermitian, second-rank tensor of odd parity~\cite{sanctuary2}.  
	
	The state operator, $\rho$, expresses the expectation values for the Hermitian observables, $A$, of~a system, and~is
	defined by the quantum trace over the operators~\cite{neumann1}:
	\begin{equation}\label{G3a}
		\left\langle A \right\rangle =\text{Tr}\left( {A}{\rho } \right)
	\end{equation} 
	Despite helicity being an element of reality, it is not observable in Minkowski space where we observe. Therefore, we express the pure state operator of Q-spin in terms of the normalized sum of the two orthogonal axes that we can observe, and include nothing about the $S^3$ hyperspace.  The state operator describes a pure state of one Q-spin,
\begin{equation} \label{stateop}
	{{\rho }}=\frac{1}{2}\left( {{I}}+ \frac{1}{\sqrt{2}}\left( {{\sigma}_{3}}+{{\sigma}_{1}} \right) \right)=\frac{1}{2}\left( {{I}}+\sigma \cdot \mathbf{r} \right)
\end{equation}
	The vector is identified $\mathbf{r}=\frac{1}{\sqrt{2}}\left( {{e}_{3}}+{{e}_{1}} \right)$ in the BFF. From~this, the~expectation values are calculated for the spin axes, $\Sigma_3, \Sigma_1$, using Equation~(\ref{stateop}),
	\begin{equation}\label{spins13}
		\begin{aligned}
			& \left\langle {{\Sigma }_{1}} \right\rangle =\left\langle {{\sigma }_{1}} \right\rangle +\underline{\underline{\underline{\varepsilon }}}\cdot \left\langle {i{\sigma }_{1}} \right\rangle =\frac{1}{\sqrt{2}}\left( {{e}_{1}}+i{{e}_{3}}Y \right) \\ 
			& \left\langle {{\Sigma }_{3}} \right\rangle =\left\langle {{\sigma }_{3}} \right\rangle +\underline{\underline{\underline{\varepsilon }}}\cdot \left\langle {i{\sigma }_{3}} \right\rangle =\frac{1}{\sqrt{2}}\left( {{e}_{3}}-i{{e}_{1}}Y \right) \\ 
		\end{aligned}
	\end{equation}
	with $\left\langle {{\sigma }_{i}} \right\rangle =+\frac{1}{\sqrt{2}}{{e}_{i}}$, and~the vector products are,
	\begin{equation}\label{31axis}
		\underline{\underline{\underline{\varepsilon }}} \cdot \left\langle i\sigma_{1} \right\rangle =+i\frac{1}{\sqrt{2}}{{e}_{3}}Y ;\text{                 } 
		\underline{\underline{\underline{\varepsilon }}} \cdot \left\langle i\sigma_{3} \right\rangle =-i\frac{1}{\sqrt{2}}{{e}_{1}}Y 
	\end{equation}
	Permuting each axis in  Equation~(\ref{spins13}) shows that the two fermionic axes are mirror states, ${{P}_{13}}\left\langle {{\Sigma }_{1}} \right\rangle ={{\left\langle {{\Sigma }_{3}} \right\rangle }^{*}}$. The first term in Equation~(\ref{spins13}) is the usual spin polarization observed. The~second shows the planes orthogonal to the axes: $e_1$ is orthogonal to $e_3Y$, and~$e_3$ is orthogonal to $e_1Y$. These terms form the wedge or~vector product from GA~\cite{GA} leading to the formulation of~helicity.
	
	In free flight, the angular momentum of the two axes, Equations~(\ref{spins13}), constructively interferes to produce the resonance spin, a boson of magnitude 1.
	\begin{equation}\label{sigma15}
		\begin{aligned}
			\Sigma _{31}={{\Sigma }_{3}}+ {{\Sigma }_{1}}
		\end{aligned}
	\end{equation}
	Substituting Equation~(\ref{spins13})  gives the free-flight boson in the BFF, $\exp \left( \pm i\frac{\pi }{4}Y \right)=\frac{1}{\sqrt{2}}\left( 1\pm iY \right)$,
	\begin{equation}\label{spinfield}
		\left\langle \mathbf{\Sigma }_{31} \right\rangle = {{e}_{1}}\exp \left( -i\frac{\pi }{4}Y \right)+{{e}_{3}}\left( +i\frac{\pi }{4}Y \right) 
	\end{equation}
	This shows each axis multiplied by a unit quaternion that rotates around the $Y$ axis. The~$e_1$ axis is rotated by  $-\frac{\pi}{4}$, and the~$e_3$ axis is rotated by $+\frac{\pi}{4}$. Hence, the two axes coincide, bisect the first quadrant, and form the resonant boson spin, labeled as $e_{31}$ in Figure~\ref{fig:Infield}.   Bisectors of all the quadrants are possible corresponding to the boson resonance spins at any instant.  Each quadrant gives the same results, so we use the first.
	
	In Figure~\ref{fig:Infield}, the~axis of linear momentum, $Y$, is orthogonal to the $\left(e_1,e_3\right)$ plane, showing once again the geometric equivalence with a head-on view of a photon with the orthogonal magnetic and electric components oscillating~out-of-phase.
	
	Equation~(\ref{spinfield}) couples the two fermionic axes, depicted in the middle figure forming the spin-1.  When a boson encounters a polarizing field, it decouples into a~fermion.

	\section{Measured~Spin}
	Rotate Equation~(\ref{spinfield}) to the LFF using Equations~(\ref{BFF}) with $\phi=0$ and contract with a polarizing field, oriented by angle $\theta_a$ in the LFF.
	\begin{equation}\label{polfield}
		\mathbf{a}=\cos {{\theta }_{a}}Z+\sin {{\theta }_{a}}X
	\end{equation}
	This gives a unit quaternion,
	\begin{equation}\label{Qspin}
		\begin{aligned}
			\mathbf{a}\cdot \left\langle {{\Sigma }_{3}} \right\rangle &= \cos \left( {{\theta }_{a}}-\theta  \right)-i\sin \left( {{\theta }_{a}}-\theta  \right)Y  \\ 
			& =\exp \left( -i\left( {{\theta }_{a}}-\theta  \right)Y \right) \\ 
		\end{aligned}
	\end{equation}
	When a boson spin encounters a field  $\mathbf{a}$, the Least Action Principle dictates that the closer axis is influenced more than the further axis. This destroys the mirror property between the axes as one aligns with the field. The~helicity stops as the second fermionic axis, orthogonal to the aligning axis, spins about the first (see the left and right panels of Figure~\ref{fig:Rqspin}). In the presence of a field, the~expectation value of the boson spin is
	\begin{equation}\label{infield}
		\mathbf{a}\cdot \left\langle \Sigma _{31} \right\rangle=  \mathbf{a}\cdot \left\langle \Sigma _{1} \right\rangle+ \mathbf{a}\cdot \left\langle \Sigma _{3} \right\rangle
	\end{equation}
	
	\subsection{Competition between~Axes}
	
	Equation~(\ref{spinfield}) shows Q-spin in its BFF. Transforming to the LFF and using Equation~(\ref{polfield}) leads to
	\begin{equation}\label{QspinBBF}
		\mathbf{a}\cdot \left\langle {{\Sigma }_{31}} \right\rangle =\frac{1}{\sqrt{2}}\left( \cos \left( {{\theta }_{a}}-\theta  \right)\exp \left( +i\frac{\pi }{4}Y \right)+\sin \left( {{\theta }_{a}}-\theta  \right)\exp \left( -i\frac{\pi }{4}Y \right) \right)
	\end{equation}
	and shows the projections determine the contributions from each axis in the LFF,
	\begin{equation}\label{proj}
		\begin{aligned}
			& \mathbf{a}\cdot {{e}_{3}}=\cos \left( {{\theta }_{a}}-\theta  \right) \\ 
			& \mathbf{a}\cdot {{e}_{1}}=\sin \left( {{\theta }_{a}}-\theta  \right) \\ 
		\end{aligned}
	\end{equation}
	and the competition between them, Equation~(\ref{QspinBBF}).
	
	Consider the angles which can be seen from Figure~\ref{fig:Infield};~using  Equation~(\ref{proj}) shows that if $\theta_a-\theta = 0$ or $\frac{\pi}{2}$, then the field is aligned with the $e_3$ or $e_1$ axis, respectively. However, polarization along these axes is reduced from unity to $\frac{1}{\sqrt{2}}$, Equation~(\ref{QspinBBF}). This is because the polarization of the orthogonal axis to the aligned axis is averaged out and, therefore, reduced.  The~angle must be aligned with the resonance boson spin to obtain full polarization.  This occurs by choosing $\theta_a-\theta = \frac{\pi}{4}$. In~Figure~\ref{fig:Infield}, this value shows the field co-linear with the bisector, $e_{31}$,  with equal contributions from both the $e_3$ and the $e_1$ axes, so the polarization has a magnitude of 1.  Choosing   $\theta_a-\theta = -\frac{\pi}{4}$ shows the field vector orthogonal to $e_{31}$ with a value of zero. 
	
	\subsection{The Boson~Projections}
	
	If the field is co-linear with the boson spin, it precesses without uncoupling with magnetic moment of $2\mu$. This is illustrated in the lower middle part of Figure~\ref{fig:Rqspin}, but~as the field is oriented further from the bisector and closer to one of the axes, the~precession changes to nutation, wobbles, and then decouples. Decoupling occurs when  the field strength increases and overpowers the boson spin-spin coupling. Alternately, as the field axis moves further from a resonance spin, and closer to one of the fermionic axes, $e_3$ or $e_1$, decoupling  occurs directly. One of the two spin axes precesses with a magnetic moment of $\mu$, while its orthogonal axis is averaged out.  
	
	In~Figure~\ref{fig:Infield}, the bisector lies $45^{\circ}$  from either the $e_3$  or $e_1$ axis.  We, therefore, assume, again based on the Least Action Principle, that within~the $45^{\circ}$ wedge on either side of the bisector, the boson spin remains intact. 	That is, the boson precesses without decoupling when the field axis lies within $22.5^{\circ}$ from the bisector. Outside the cone, decoupling occurs, and one of the two axes, $e_3$  or $e_1$, aligns. 
	Depending on these orientation effects, Q-spin either persists as a boson or~rapidly decouples to a~fermion.  
	
	Working in the first quadrant defined by $(e_1,e_3)$, and using Equation~(\ref{BFF}) with $\phi=0$, the relation between the bisector in the BFF and the LFF is
	\begin{equation}\label{ZX}    
		\begin{aligned}
			& \left( {{e}_{3}}+{{e}_{1}} \right)=  \left( \cos \theta -\sin \theta  \right)Z+\left( \cos \theta +\sin \theta  \right)X \\
			& \left( {{e}_{3}}-{{e}_{1}} \right)=\left( \cos \theta +\sin \theta  \right)Z+\left( \sin \theta -\cos \theta  \right)X\\
		\end{aligned}
	\end{equation}
	Alternately, Equation~(\ref{proj}) projects $e_1$ and $e_3$ onto the field direction. First, we express Q-spin in its BFF (see Equation~(\ref{spins13})).
	\begin{equation}\label{ZX1} 	
		\left\langle \Sigma _{31} \right\rangle =\left\langle {{\Sigma }_{3}} \right\rangle +\left\langle {{\Sigma }_{1}} \right\rangle  =\frac{1}{\sqrt{2}}\left(\left( {{e}_{3}}+{{e}_{1}}\right)+i\left( {{e}_{3}}-{{e}_{1}} \right)Y \right)
	\end{equation}
	Substitution of Equation~(\ref{ZX}) leads to
	\begin{equation}\label{qspinlff}
		\begin{aligned}
			\mathbf{a}\cdot \left\langle \Sigma _{31} \right\rangle
			&=\frac{1}{\sqrt{2}}\left( \left( \cos \theta -\sin \theta  \right){{\operatorname{e}}^{-i{{\theta }_{a}}Y}}+\left( \cos \theta +\sin \theta  \right){{\operatorname{e}}^{+i\left( \frac{\pi }{2}-{{\theta }_{a}} \right)Y}} \right) \\ 
			&=\frac{1}{\sqrt{2}}\left( \left( \cos {{\theta }_{a}}+\sin {{\theta }_{a}} \right){{\operatorname{e}}^{i\theta Y}}-\left( \cos {{\theta }_{a}}-\sin {{\theta }_{a}} \right){{\operatorname{e}}^{-i\left( \frac{\pi }{2}-\theta  \right)Y}} \right) 
		\end{aligned}
	\end{equation}
	Each axis is multiplied by a unit quaternion, and the two are orthogonal. In~Equation~(\ref{ZX}), and~contracting with $Z$ and $X$, we see that the first term is the projection of the bisectors along the LFF $Z$ axis, and~the second term is the projection along $X$ axis. These projections depend only on the spin's orientation via $\theta$,
	\begin{equation}
		\begin{aligned}
			& \left( {{e}_{3}}+{{e}_{1}} \right)\cdot Z=\left( \cos \theta -\sin \theta  \right) \\ 
			& \left( {{e}_{3}}+{{e}_{1}} \right)\cdot X=\left( \cos \theta +\sin \theta  \right) 
		\end{aligned}
	\end{equation}
	A similar interpretation follows from the second line of Equation~(\ref{qspinlff}). These can be compared to Equation~(\ref{QspinBBF}), which projects the axes $e_3$ and $e_1$ rather than their bisector, as shown in Figure~\ref{fig:Infield}. Either equation can be used to determine the spin-polarization.
	
	\subsection{Q-Spin as~Quaternions}
	
	Equations~(\ref{QspinBBF}) and (\ref{qspinlff}) lead to a quaternion in terms of the angle differences  $\left(\theta_{31}-\theta_a\right)$, which is independent of $\theta$.
	\begin{equation}\label{theta21}
		\mathbf{a}\cdot \left\langle \Sigma _{31} \right\rangle =\exp \left( i\left( {{\theta }_{31}}-{{\theta }_{a}} \right)Y \right)
	\end{equation}
	In the first quadrant, the bisector is normalized to
	\begin{equation}\label{e31}
		{{e}_{31}}=\frac{1}{\sqrt{2}}\left( {{e}_{3}}+{{e}_{1}} \right)
	\end{equation}
	with the angle given by
	\begin{equation}\label{e31}
		\begin{aligned}
			& {{e}_{31}}\cdot Z=\cos {{\theta }_{31}} \\ 
			& {{e}_{31}}\cdot X=\sin {{\theta }_{31}} \\ 
		\end{aligned}
	\end{equation}
	Clearly, Equation~(\ref{theta21}) shows that the boson aligns with the field when the angles are equal, $\theta_a = \theta_{13}$.  As discussed above, when aligned and when a small amount offsets the field, the boson does not decouple but precesses as a spin-1 with a magnetic moment of $2\mu$. This is shown in the middle lower panel of Figure~\ref{fig:Rqspin}. 
	
	To determine which fermion axis will align, we use Equation ({\ref{QspinBBF}}) or Equation (\ref{qspinlff}) and determine which has the larger magnitude. The larger axis aligns, and its sign then determines if the aligned spin is up or down. Note that the two axes, $e_3$ and $e_1$, have opposite magnetic moments of $\mu$ each. Consider further Equation~(\ref{QspinBBF}), which can be written as,
	\begin{equation}\label{Qgeneral}
		\begin{aligned}
			\mathbf{a}\cdot \left\langle \Sigma _{31} \right\rangle & =\exp \left( i\left( \frac{\pi }{4}-\left( {{\theta }_{a}}-\theta  \right) \right)Y \right) \\ 
			& ={{e}^{i\frac{\pi }{4}Y}}{{\operatorname{e}}^{+i{{\theta }}Y}}{{\operatorname{e}}^{-i\theta_{a} Y}} \\ 
		\end{aligned}
	\end{equation}
	This shows that the product of three quaternions determines Q-spin. The~first is a phase, Equation~(\ref{31axis}), that maintains Alice and Bob's spins anti-parallel; the second is a geometric factor that orients the disc and is determined at the source; the third is a field quaternion that pulls the~axis.
	\section{The EPR~Paradox}
	These features of Q-spin are crucial~\cite{sanctuary2} in understanding the extra correlation found in coincidence EPR experiments~\cite{clauser,aspect1,aspect2,zeilinger} and which is more fully discussed in paper three~\cite{sanctuary2}.  That is, the~spin orientation relative to the field direction and strength of spin coupling relative to the field strength provide two mechanistic pathways for boson decoupling.  The~violation of BI~\cite{Bell64} is due to the transition from a free-flight boson to a measured~fermion.
	
	Even though measurement reveals two real states, spin is a complex element of reality and is defined by Equation~(\ref{sigmar}). We have shown down to Equation~(\ref{Qgeneral}) that unit~quaternions govern spin.  
	
	Consider the correlation between an EPR pair, which, using Q-spin, is written as a product state between Alice and Bob.
	\begin{equation}\label{fullcorr1}
		\begin{aligned}
			& E\left( a,b \right)=\mathbf{a}\cdot \frac{1}{2}\left(\left\langle \Sigma^A _{31} \right\rangle  {{\left\langle \Sigma^B _{31} \right\rangle }^{*}}+{{\left\langle \Sigma^A _{31} \right\rangle }^{*}} \left\langle \Sigma^B _{31} \right\rangle\right) \cdot \mathbf{b} \\ 
			& =\frac{1}{2}\exp \left( i\left( \frac{\pi }{2}-\left( {{\theta }_{a}}-\theta  \right) \right)Y \right)\exp \left( i\left( \frac{\pi }{2}+\left( {{\theta }_{b}}-\theta  \right) \right)Y \right)+c.c. \\ 
			& =-\cos \left( {{\theta }_{a}}-{{\theta }_{b}} \right) \\ 
		\end{aligned}
	\end{equation}
	We have taken the angle $\theta$ to be $\theta \pm \pi$ for Alice and Bob, so the two spins at the source have a common orientation of $\theta$ that differs by $\pi$ to keep the two anti-parallel.  Since we only measure in real space, similar to light, the complex part can be removed by forming linear or circularly polarized components.  
	
	In EPR coincidence experiments, only the real part is measured, but~the complexity is essential to give the observed result, $-\cos(\theta_a - \theta_b)$, with~a violation of BI \cite{clauser} of CHSH =2$\sqrt{2}$. However, to obtain this result, both spins of Alice and Bob must be complex,  Equation~(\ref{fullcorr1}). If one spin is polarized, there is no helicity, and only the scalar part of the quaternion is present.
	\begin{equation}\label{noHelicity}
		\begin{aligned}
			\exp \left( i\left( \frac{\pi }{2}+\left( {{\theta }_{b}}-\theta  \right) \right)Y \right)		& ={{e}^{i\frac{\pi }{2}Y}}\exp \left( i\left( {{\theta }_{b}}-\theta  \right)Y \right)\\
			& \xrightarrow{\text{no helicity}}i \cos \left( {{\theta }_{b}}-\theta  \right)
		\end{aligned}
	\end{equation}
	We separate the $\frac{\pi}{2}$ phase needed for anti-correlation. Only the product state survives even if one spin is coherent,
	\begin{equation}\label{prodcorreoation}
		\begin{aligned}
			E\left( a,b \right)=
			& -\frac{1}{2}\exp \left( i\left( {{\theta }_{a}}-\theta  \right)Y \right)\cos \left( {{\theta }_{b}}-\theta  \right)+c.c. \\ 
			& =-\cos\left({{\theta }_{a}}-\theta\right) \cos\left({{\theta }_{b}}-\theta\right)  \\ 
		\end{aligned}
	\end{equation}
Equation~(\ref{fullcorr1}) shows Alice and Bob's particles must both be boson spins to obtain the full correlation. If one or the other has decoupled into a fermion, only a product state is possible. The correlation from coherence is only present when Alice and Bob spin simultaneously and precess as a spin-1.
	
	The product state, Equation~(\ref{prodcorreoation}),  satisfies BI with CHSH = 2, whereas the full correlation in Equation~(\ref{fullcorr1}) with CHSH = 2$\sqrt{2}$ violates the inequality. The part that violates BI is the correlation due to helicity~\cite{sanctuary2,sanctuary3}. Helicity replaces non-locality. Bell's Theorem~\cite{BellBert} proves that real classical systems cannot violate his inequality, but is inapplicable to complex complementary quantum properties that exist is two distinct convex spaces.
	
	\section{Discussion}
	
	In free flight, the spinning disc is reminiscent of the worldsheet Susskind introduced~\cite{susskind}. A~ spin system is also an anyon~\cite{Wilczek}, which can be either a fermion or a boson. An important point about a boson in free flight is that the spinning axis averages out the boson polarization, Figure~\ref{fig:freespin}, so only the odd-parity helicity remains. Upon measurement, a transition from a boson to a fermion occurs, giving the usual two-state Dirac~spin. 
	
	The motivation behind Twistor theory, \cite{penrose1,penrose2}, is Nature is fundamentally complex, and we measure the real part. Q-spin supports this concept, as seen from the discussion above on EPR. The coherence carried by the helicity accounts for the violation of BI~\cite{sanctuary3}.  Dropping the complexity removes the quantum coherence, leaving only the classical correlation that obeys BI. Without~helicity, we have Dirac's two-state spin and no coherent~properties. 
	
	Introducing the bivector into spin algebra significantly changes our view of a spin from a structureless point particle of intrinsic angular momentum in Minkowski space to a four-dimensional structured spin with extrinsic angular momentum in spin spacetime. Four axes compose Q-spin. One is the axis of linear momentum spun by the quaternion. Two more are the magnetic axes, which couple to give the fourth, being the boson spin.  Figure~\ref{fig:freespin}.
	
	The question arises: does Q-spin exists and is it more fundamental than point-particle Dirac spin.  That Q-spin and a photon have structure and properties in common is compelling, Figure~\ref{fig:freespin}. Other quantum observables come in complementary pairs, like position and momentum, $etc.$, in~spaces that are the inverse of each other. It is, therefore, reasonable that spin also has two complementary properties, real polarization in its spin spacetime and~imaginary coherence on the $S^3$ hypersphere.  
	
	Measurement has a central premise that the act of observation perturbs the system. Q-spin makes a distinction between the measurement of a fermi electron, $e^-_F$ (polarized, particle, fermion, even to parity) and the free flight of a boson electron, $e^-_B$  (coherence, wave, boson, odd to parity). They epitomize particle-wave duality.   An advantage of Q-spin lies in its expression in terms of quaternions.  One can envisage a spin as a stable qubit where the two axes carry opposite spin. The~evolution is calculated using the products of quaternions, Equation~(\ref{Qgeneral}), and~the coherence maintains correlation between~gates. 
	
	The mathematical foundations of Q-spin are the same as those of Dirac spin.  Changing the symmetry from SU(2) to $Q_8$ is our only modification to the Dirac field. The solution to the 2D Dirac equation and the spin spacetime gamma algebra carry over from the usual treatment without difficulty. One advantage of Q-spin is that it gives alternate interpretations of some troubling properties: non-locality is repudiated~\cite{sanctuary3}; negative energies of the antimatter particle are resolved; and several other changes challenge our existing view of the microscopic, \cite{sanctuary4}.  
	
	Nature is complex, and a free-flight electron is a boson of odd parity, and a measured electron is a fermion of even~parity.

	\vspace{6pt}
	\funding{{This research received no external funding.}} 
	
	\dataavailability{{No new data were created or analyzed in this study. Data sharing is not applicable to this article.}} 
	
	\acknowledgments{The author is grateful to {Hillary Sanctuary},  EPFL Switzerland, for engaging in helpful~discussions.}
	
	\conflictsofinterest{{The author declares no conflicts of interest.}} 
	
	\begin{adjustwidth}{-\extralength}{0cm}
		
		\reftitle{References}

		\PublishersNote{}
	\end{adjustwidth}
\end{document}